\documentclass[preprint, pre,aps,showpacs,showkeys]{revtex4}

\usepackage{graphicx}
\begin{document}
\title{
Self-focusing and envelope pulse generation in nonlinear 
magnetic metamaterials
}

\author{
I. Kourakis$^{1 }$,
N. Lazarides$^{2, 3}$ and G. P. Tsironis$^{2, 4}$
}
\affiliation{
$^{1}$ Institut f\"ur Theoretische Physik IV,
Fakult\"at f\"ur Physik und Astronomie, 
Ruhr--Universit\"at Bochum, D-44780 Bochum, Germany \\
$^{2}$Department of Physics, University of Crete,
and
Institute of Electronic Structure and Laser,
FORTH,
P. O. Box 2208, 71003 Heraklion,  Greece \\
$^{3}$ 
Department of Electrical Engineering,
Technological Educational Institute of Crete,
P. O. Box 140, Stavromenos, 71500, Heraklion, Crete, Greece \\
$^{4}$
Facultat de Fisica, Department d'Estructura i Constituents de la Materia,
Universitat de Barcelona, Av. Diagonal 647, E-08028 Barcelona, Spain
}

\date{Submitted \today}

\begin{abstract}
The self-modulation of waves propagating in nonlinear 
magnetic metamaterials is investigated.
Considering the propagation of a modulated amplitude magnetic field
in such a medium, we show that the self-modulation of the carrier
wave leads to a spontaneous energy localization via the generation of 
localized envelope structures (envelope solitons),
whose form and properties are discussed.
These results are also supported by numerical calculations.
\end{abstract}

\maketitle
Increased research efforts have recently focused on \emph{metamaterials}
i.e., artificially structured materials that exhibit electromagnetic (EM)
properties not available in naturally occuring materials. 
For example, arrays of subwavelength split-ring resonators (SRRs) have been
utilized, as proposed by Pendry {\it et al.} \cite{Pendry1-2}, to create magnetic
metamaterials (MMs) with negative magnetic permeability $\mu$ up to 
Terahertz (THz) frequencies \cite{Yen,Katsarakis}.
There are only few natural materials responding magnetically at these
frequencies. However, the magnetic effects in these materials are typically
weak and often exhibit narrow bands (see Ref. \cite{Yen} and refs. therein).
Thus, the realization of MMs at such frequencies will affect 
THz optics substantially, while it promises new device applications. 
Moreover, MMs with negative $\mu$ can be combined with 
plasmonic wires that exhibit a negative permittivity $\epsilon$ \cite{Pendry}, 
producing metamaterials which possess simultaneously negative values of $\mu$ and $\epsilon$ 
\cite{Pendry3-Ramakrishna}.
The behavior of these metamaterials obeys a negative value of the refraction index,
so they are usually referred to as negative index media or left-handed
materials (LHMs).

A concise analytical framework for the nonlinear behavior of LHMs was only quite 
recently proposed \cite{Zharov-OBrien}, by considering the SRRs embedded in 
a Kerr-type medium, leading to a dependence of the effective values
of $\epsilon$ and $\mu$ on the EM field amplitude(s) of a propagating wave,
roughly analogous to the Kerr effect in nonlinear optics.
The SRR is modeled as a nonlinear RLC
circuit, featuring an Ohmic resistance $R$, a self-inductance $L$, and 
a nonlinear capacitance $C$ (due to the nonlinear dielectric filling its slit).
Relying on this formulation, a nonlinear model for propagating modulated EM waves 
in LHMs predicts the possibility for modulational instability of EM
waves and the occurrence of envelope bi-solitons, in fact exact solutions of a set of 
coupled nonlinear Schr\"odinger equations \cite{NL-IK}. 
The role of the anomalous self-steepening effect in modulational instability
in LHMs has been studied by Wen {\it et al.}\cite{Wen}.
The spontaneous generation of a magnetic field due to the ponderomotive interaction 
of high-frequency EM fields with low-frequency electrostatic background 
oscillations in the medium (plasmons) was suggested in Ref. \cite{IK2}.
A one-dimensional (1D) discrete array of nonlinear SRRs was considered 
in Ref. \cite{NL2}, which supports highly localized excitations in the form 
of discrete breathers.
This model incorporates the nonlinearity of SRRs and the coupling
among them at neighboring lattice sites via their mutual inductance $M$,
thus accounting for Fourier dispersion of propagating charge waves. 
The combination of a nonlinear and dispersive lattice behavior allows one to anticipate 
the formation of nonlinear localized structures (solitons), 
eventually sustained by a balance among these mechanisms.
In this Brief Report, we consider the propagation of a
weakly nonlinear charge variation in a nonlinear MM.
Considering the self-interaction of the charge carrier wave, 
we show that the modulated wavepacket may be unstable, and that it may evolve towards 
the formation of localized structures in the form of envelope solitons.

We consider an EM wave propagating in a 1D array of $N$ identical SRRs oriented 
in a direction perpendicular to the principal lattice axis,
forming thus a magnetic-crystal-like arrangement with lattice spacing $D$. 
In the following we adopt the description (and notation) in Ref. \cite{NL2}, 
thus summarizing the essential building blocks of the theory in a self-contained manner,
yet omitting unnecessary details.
Considering a Kerr-type dependence of the permittivity $\epsilon$ 
of the dielectric filling the SRR slits
on the wave electric field $\mathbf{E}$, 
viz. $\epsilon = \epsilon_0 (\epsilon_l + \alpha |\mathbf{E}|^2/E_c^2)$,
where $\epsilon_0$ and $\epsilon_l$ respectively denote the
permittivity of the vacuum and the linear part of the
permittivity ($E_c$ is a characteristic electric field constant; 
$\alpha = +1/-1$ accounts for a focusing/defocusing nonlinearity),
the electric charge $Q_n$ stored in the $n$th capacitor is given by 
$Q_n = C_l U_n [1 +  \alpha U_n^2/(3 \epsilon_l U_c^2)]$, 
where $C_l$ is the linear capacitance of the SRRs, 
$U_n = d_g E_{g, n}$ is the voltage across the slit of the $n$th SRR
(viz. $\mathbf{E}_{g}$ is the electric field induced along the slit) 
and $U_c = d_g E_c$. 

Combining the above considerations, 
and approximating the nonlinearity by the first nonlinear term
in a Taylor expansion of its precise form
(i.e., keeping up to cubic terms in equation (6) in Ref. \cite{NL2}),
the charge stored in the $n$th capacitor is found
to obey the evolution equation
\begin{equation}
   \frac{d^2}{d \tilde{t}^2} (\lambda q_{n-1} - q_n + \lambda q_{n+1}) - q_n 
   + \frac{\alpha}{3 \epsilon_l} q_n^3  =  \gamma
   \frac{d q_{n}}{d \tilde{t}} - \varepsilon(\tilde{t}) \, ,
\label{eq}
\end{equation}
where $q_n = Q_n/(C_l d_g E_c)$ denotes the reduced charge, 
$\tilde{t} = t/\omega_l \equiv t (L C_l)^{1/2}$, 
and $\lambda = M/L$ is the inter-site coupling constant. 
In the right-hand side (rhs),
$\gamma = R C \omega_l$ accounts for the Ohmic impedance (friction) force 
$F_{imp} = \gamma I_n/(U_c \omega_l C_l)$ ($I_n = d Q_n/dt$) and 
$\varepsilon = {\cal E}/U_c$ is related to the electromotive force
${\cal E}$ induced in each SRR due to the applied EM field.
The rhs will be omitted in the following, i.e., by setting 
$\gamma = \varepsilon = 0$, thus namely neglecting Ohmic losses and 
electromotive forcing; these effects may be included a posteriori 
(e.g., as perturbations to the analytical solutions anticipated here) in future work.
In what follows, $k$ and $\omega$ represent their corresponding normalized 
values  $k \rightarrow k\, D$, and $ \omega  \rightarrow \omega /\omega_l$.
Then, the linear dispersion relation (DR) can be written in 
the form
$\omega = [1 - 2 \lambda \cos(k)]^{-1/2}$, governing the propagation
of magnetoinductive waves in such systems\cite{Sydoruk-Shamonina}. 
This DR possesses a finite cutoff at 
$\omega_{max} = \omega(k=0) = 1/\sqrt{1 - 2 \lambda}$, 
and describes an \emph{inverse} optic law, 
i.e., the group velocity $v_g \equiv \omega'(k) =-\lambda \omega^3 \sin k$
is negative for all $k'$s within the first Brillouin zone, viz. $k \in [0, \pi]$.
The wavepacket envelope therefore propagates (at the group velocity $v_g$) 
in the \emph{opposite} direction with respect to the carrier wave 
(propagating at the phase speed $v_{ph} = \omega/k$). 
The frequency band is therefore bounded by 
$\omega_{min} =\omega(k=\pi) = (1 + 2 \lambda)^{-1/2}$ and 
$\omega_{max}$. 
Note that $\omega_{max}$ is increased by increasing the value of $\lambda$, 
while $\omega_{min}$ is decreased in the same time. 
These results are apparently valid for $\lambda \le 1/2$, 
the physically meaningfull range for the system in consideration.
A nonlinear generalization of the DR is obtained via a rotating 
wave approximation, 
i.e., by substituting $q_n = \hat q \exp[i (k n - \omega \tilde{t} )] +$ c.c. in 
the \emph{nonlinear} form of Eq. (\ref{eq}) and retaining only first
order harmonics. One thus obtains
\begin{equation}
    \omega^2(k; |\hat q|^2) 
    = \left({1 - {\alpha} |\hat q|^2} / {\epsilon_l}\right)
      \left({1 - 2 \lambda \cos(k)}\right)^{-1}  , 
\label{NLDR}
\end{equation}
which incorporates the amplitude-dependence of the wave frequency. 
Assuming this dependence to be weak, and considering a modulated wave frequency $\omega$ 
and wavenumber $k$ close to the carrier values $\omega_0$ and $k_0$, respectively,
one may expand  as 
\begin{eqnarray}
   \omega - \omega_0 & \approx & 
     \left. \frac{\partial \omega}{\partial k}\right|_{\omega=\omega_0} (k-k_0) 
       + \left. \frac{1}{2} \frac{\partial^2 \omega}{\partial k^2}\right|_{\omega=\omega_0}
       (k-k_0)^2 
+
      \frac{\partial \omega(k)}{\partial |\hat q|^2}\biggr|_{\hat q= \hat q_0}  
             (|\hat q|^2 - |\hat q_0|^2) \, ,
\label{expansion}
\end{eqnarray}
where $\hat q_0$ is a reference (harmonic wave, constant) amplitude. 
Considering slow space and time variables $X$ and $T$, and thus setting 
$\omega - \omega_0 \rightarrow i \partial/\partial T$ and 
$k - k_0 \rightarrow - i \partial/\partial X$, 
one readily  obtains the nonlinear Schr\"odinger- (NLS-)type equation
\begin{eqnarray}
  i \biggl( \frac{\partial \psi}{\partial T} + v_g \frac{\partial \psi}{\partial X} \biggr) 
  + P \frac{\partial^2 \psi}{\partial X^2} + Q  (|\psi|^2 - |\psi_0|^2) \psi = 0\, ,
\label{NLSE0}
\end{eqnarray}
where we have set 
$\psi = \hat q$ and $\psi_0 = \hat q_0$, and defined the dispersion coefficient 
$P \equiv {\omega''(k)}/{2}$, so that
\begin{equation}
\label{Pcoeff}
  P = -\lambda \omega^5 
       (\lambda \cos^2 k + \cos k  - 3 \lambda) /2 \, ,
\end{equation}
and the nonlinearity coefficient 
$Q=-(\partial \omega/\partial |\psi|^2\bigr|_{\psi = \psi_0} $, viz.
\begin{equation}
   Q = \frac{\omega}{2} \frac{\alpha}{\epsilon_l} 
   \frac{1}{1 - \frac{\alpha}{\epsilon_l} |\psi_0|^2} 
   \approx \frac{\omega}{2} \frac{\alpha}{\epsilon_l} \, .
\label{Qcoeff}
\end{equation}
Note that we have assumed $|\psi_0|\ll 1$, and thus neglected the dependence
on $|\psi_0|$ everywhere. Upon a Galilean transformation, viz. 
$\{X, T\} \rightarrow \{X - v_g T, T\} \equiv \{\zeta, \tau\}$,
and a phase shift $\psi \rightarrow \psi e^{-i Q  |\psi_0|^2 \tau}$, 
one obtains the usual form of the NLS equation
\begin{eqnarray}
     i \frac{\partial \psi}{\partial \tau} + P \frac{\partial^2 \psi}{\partial \zeta^2} 
        + Q  |\psi|^2 \psi = 0\, ,
\label{NLSE}
\end{eqnarray}
which is known to occur in a variety of physical contexts \cite{Daumont,Dauxois}. 
Eq. (\ref{NLSE}), along with the expressions (\ref{Pcoeff}) and (\ref{Qcoeff}), 
are the strong result of this calculation, to be retained in the analysis which follows.

The evolution of a modulated wave whose amplitude is described 
by Eq. (\ref{NLSE}) essentially depends on the sign of the coefficients 
$P$ and $Q$ \cite{Dauxois}. 
In specific, if $PQ < 0$ the wavepacket is modulationally stable
(and may propagate in the form of a dark-type envelope, 
i.e., a \emph{hole} envelope soliton), while for  $PQ > 0$ the wavepacket is modulationally 
\emph{unstable}. 
In the latter case, the wave may respond to random external perturbations 
(noise) by breaking-up to a ``sea'' of erratic oscillations (collapse) or 
(as suggested by analytical and  numerical investigations \cite{Daumont,Dauxois,Peyrard})
by localizing its energy via the formation of a series of localized envelope structures,
i.e., propagating wavepackets modulated by a pulse-shaped envelope. 
We find that $P$ is negative for low $k$, while it changes sign at some critical value 
$k_{cr}= \cos^{-1} \bigl[ ( -1 + \sqrt{1 + 12 \lambda^2})/2 \lambda \bigr]$, 
thus acquiring positive values for $k > k_{cr}$. 
On the other hand, the sign of $Q$ is simply determined by the nature of the 
nonlinearity, i.e., $Q$ is positive (negative) for $\alpha = $ +1 (-1). 
We may therefore distinguish two cases.
For $\alpha = +1$, 
the wave is modulationally stable ($P Q < 0$) for $k <k_{cr}$,
while it is unstable ($P Q > 0$) for $k > k_{cr}$. 
For $\alpha = -1$, the wave is modulationally 
unstable ($P Q > 0$) for $k<k_{cr}$, 
while it is stable ($P Q < 0$) for $k > k_{cr}$. 

To see this, first check that Eq.
(\ref{NLSE}) supports the plane wave solution 
$\psi = \psi_0 \exp(i Q |\psi_0|^2 T)$;
the standard linear analysis
consists in perturbing the amplitude by setting: 
${\hat \psi} \, =
\, {\hat \psi}_0 \, + \, \delta \, {\hat \psi}_{1, 0}
\cos{({\tilde k} \zeta - {\tilde \omega} \tau)}$ 
(the perturbation
wavenumber $\tilde k$ and the frequency $\tilde \omega$ are
distinguished from the  carrier wave quantities, $k$ and $\omega$). 
One thus obtains the perturbation DR
$\tilde \omega^2  =  P \, \tilde k^2 \, (P
\tilde k^2 \, - \, 2 {Q} |\hat \psi_{1, 0}|^2 )$.
One immediately sees that if $P Q < 0$, the
amplitude $\psi$ will be \emph{stable} to external perturbations.
On the other hand, if $P Q > 0$,
the amplitude $\psi$ is \emph{unstable} for $\tilde k <
\sqrt{2 {Q}/{P}}  |\hat \psi_{1, 0}|$; i.e. for perturbation
wavelengths larger than a critical value. The maximum perturbation growth rate 
is then found to be $\sigma_{max} = |Q| |\hat \psi_{1, 0}|^2$, 
and will therefore be inversely proportional to both 
$\epsilon_l$ and $E_c^2$ 
[as can be seen by recovering dimensions in Eq. (\ref{NLSE})]. 
This \emph{modulational instability} mechanism is tantamount to the
Benjamin-Feir instability in hydrodynamics,
also long-known as an energy localization mechanism in solid state
physics and nonlinear optics, among other physical contexts \cite{Dauxois}.
This type of analysis  allows for a numerical investigation of the
stability profile in terms of intrinsic medium properties.

Eq. (\ref{NLSE}) is integrable and possesses a number of exact solutions.
Of particular interest to us are its constant profile, 
localized envelope solutions of the bright- (dark-) type, 
obtained for a positive (negative) value of $P Q$,
which are of the form $\psi= \psi_0 \exp(i \Theta)$. 
The bright-type soliton solutions
are given by \cite{FS}
\begin{eqnarray} 
  \psi_0 & =& 
  \label{brightsol}
   \psi_0' {\rm{sech}} \biggl( \frac{\zeta - v_e \, \tau}{L} \biggr) , \\
  \label{brightsol1}
  \Theta & =& [ v_e  \zeta + (\Omega -
  {v_e^2}/{2} ) \tau ] / {2 P} ,
\end{eqnarray}
where $v_e$ is the envelope velocity; $L$ and $\Omega$ represent the
pulse's spatial width and oscillation frequency (at rest),
respectively, and $L \psi_0' = (2P/Q)^{1/2}$.
Note that  $\psi_0'$ is independent of the pulse (envelope)
velocity $v_e$ here.

A \emph{dark} (\emph{black} or \emph{grey}) envelope wavepacket,
looks like a propagating localized \emph{hole} (a \emph{void})
amidst a uniform wave energy region. The exact
expression for \emph{black} envelopes reads \cite{FS}:
\begin{eqnarray}
\label{darksoliton}
\psi_0 &=& {\psi'}_0 \, \biggl| {\rm{tanh}} \biggl( \frac{\zeta - v_e \,
\tau}{L} \biggr) \biggr| ,  \\   
\label{darksoliton1}
\Theta &=&
[ v_e \zeta + ( 2 P Q {{\psi'}_0}^2 - {v_e^2}/{2} )
\tau ] / {2 P}  , 
\end{eqnarray}
where ${\psi'}_0 L = (2|P/Q|)^{1/2}$.
The \emph{grey}-type envelope is given by \cite{FS}
\begin{eqnarray} 
\label{greysoliton}
  \psi_0  &= &{\psi'}_0  \biggl[ 1 - d^2 \,  {\rm{sech}}^2
  \biggl( \frac{\xi}{L} \biggr) \biggr]^{1/2} , \\ 
\label{greysoliton1}
  \Theta &=&  [ V_0\,\zeta 
  -( V_0^2 /2 - 2 P Q {\psi'}_0^2 ) \tau +
  \Theta_{0} ] / {2 P} 
  -S \sin^{-1} \left\{ 
    {d\, \tanh\biggl(\frac{\xi}{L} \biggr)}
  {\biggr[  1 - d^2\, {\rm{sech}}^2
\biggl(\frac{\xi}{L} \biggr) \biggr]^{-1/2}}
\right\} .  
\end{eqnarray}
Here, $\Theta_{0}$ is a constant phase, $\xi=\zeta - v_e \tau$, and  
$S = {\rm{sign}} (P) \times {\rm{sign}} (v_e -V_0)$.
The pulse width
$L = |P/Q|^{1/2}/d \psi_0'$ now also depends on the real
parameter $d$, given by: 
$ d^2 = 1 + (v_e - V_0)^2/({2 P Q} {{\psi'}_0^2})  \le  1 $.
The (real) velocity parameter $V_0$ satisfies the relation
$V_0 - \sqrt{2 |P Q|\, {\psi'}_0^2}  \le  v_e 
\le V_0 + \sqrt{2 |P Q|\, {\psi'}_0^2}$ .
For $d = 1$  one recovers the {\em black} envelope soliton.

We performed numerical simulations using Eq. (\ref{eq}) with the 
complete form of the nonlinearity \cite{NL2} and initial conditions
$q_n = \psi_0 \, \cos(n\,k-\omega \tilde{t})$,
with $\psi_0$ in the form given in Eq. (\ref{brightsol}) and Eq. (\ref{darksoliton})
for bright and dark envelope solitons, respectively.
The results are illustrated in Figs. 2 and 3 for both focusing and 
defocusing nonlinearity ($\alpha= +1$ and $-1$), along with the 
corresponding analytic expressions, 
Eqs. (\ref{brightsol}) and (\ref{darksoliton}).
In Fig. 3, we have also depicted the envelope of the grey-type soliton,
given by Eq. (\ref{greysoliton}) with large "greyness", i.e., with $d=0.995$.
The numerically obtained envelopes are shown at some instant after they 
have performed 
at least ten revolutions around the lattice, that is, more than $4200$ 
time units (t.u.).  However, they seem to be stable for 
much longer time intervals (at least up to $5\times 10^4$ t.u. for the cases we checked).
This looks surprising, given that for the value of $\lambda$ used in the 
calculations ($0.2$ to be close to realistic values appearing in real systems)
the system is still far from being continuous. This fact, along with 
the approximation of the nonlinear term, makes the expressions for the 
envelopes presented above to be the only approximate solution. 
Thus, they induce radiation in the lattice which eventually deforms the 
envelopes. Apparently, the simulations based on the discrete model show fairly 
good agreement with the analytical expressions derived from the continuous 
approximation of that model. In any case, the analytical expressions seem to 
reproduce, at least qualitatively,  the observed envelopes.
We should note that our numerical simulation also provides evidence for
dynamical dark soliton formation via modulational instability of a
slightly modulated plane wave. This is a strongly nonlinear stage of the
wave amplitude's evolution, which is not predicted by the linear
amplitude perturbation theory employed above. The same procedure for
bright solitons results in erratic oscillations.

In conclusion, we have shown that the propagation of a modulated
EM wave packet in a nonlinear MM, in the form of a lattice of SRRs,
is characterized by amplitude modulation due to carrier wave self-interaction. 
An EM wave packet may be modulationally stable, and then propagate in the form
of a localized void (a hole, amidst a uniform charge density),
or it may be intrinsically unstable,
and thus possibly evolve towards the formation of envelope pulses (bright solitons). 
Explicit expressions for these nonlinear magnetoinductive excitations, 
whose existence is supported by
numerical calculations, are provided in terms of the intrinsic material parameters. 
These results are of relevance in metamaterial-related applications, 
in materials which may be ``tuned'' appropriately in order for the forementioned 
excitations to occur, pretty much like optical fiber solitons in nonlinear optics. 

{\it Acknowledgments.} 
Support from the FWO (Fonds Wetenschappelijk
Onderzoek-Vlaanderen, Flemish Research Fund) during the course of
this work is gratefully acknowledged.
IK's work was carried out during a short-term Research Associate
appointment at the University of Gent, Belgium. 
Prof. F. Verheest is herewith warmly thanked for actively
encouraging this research visit.
IK's gratitude and appreciation goes to Dr. T.
Cattaert, for long scientific discussions and friendly support.
GPT and NL  acknowledge  support from the grant "PYTHAGORAS II" 
(KA. 2102/TDY 25)
of the Greek Ministry of Education and the European Union, 
and grant 2006PIV10007 of the Generalitat de Catalunia.

\newpage 

\textbf{Figure captions}

\bigskip 

\textbf{Figure 1}

(a) The linear dispersion $\omega=\omega (k)$, 
 and (b) the dispersion coefficient $P=P (k)$, 
 for $\lambda=0.05$ (black-solid curve), $0.2$ (red-dotted curve),
 $0.35$ (green-dashed curve).

\bigskip 

\textbf{Figure 2}

Dark-type modulated wavepackets, for $\lambda=0.20$,
 $\epsilon_{\ell} = 2$, $\psi_0' =0.1$, $N=100$, and 
 (a) $\alpha=-1$, $k=1.12 > k_{cr}$ ($\omega=1.10$), $L=3.74$;  
 (b) $\alpha=+1$, $k=0.86 < k_{cr}$ ($\omega=1.16$), $L=4.51$.
 The filled circles correspond to the $q_n$'s, with $q_n$ the charge at 
 site number $n$,
 while the black-dotted curve serves as a guide to the eye.
 The red-solid curves are the envelope Eq. (\ref{darksoliton}), and the 
 green-dashed curves the envelope Eq. (\ref{greysoliton}), with parameters
 as in (a) and (b), and $d=0.995$.

\bigskip 

\textbf{Figure 3}

Bright-type modulated wavepackets, for $\lambda=0.20$,
 $\epsilon_{\ell} = 2$, $\psi_0' =0.1$, $N=100$, and 
 (a) $\alpha=-1$, $k=0.86 < k_{cr}$ ($\omega=1.16$), $L=9.02$;
 (b) $\alpha=+1$, $k=1.12 > k_{cr}$ ($\omega=1.10$), $L=7.48$.
 The filled circles correspond to the $q_n$'s, with $q_n$ the charge at 
 site number $n$,
 while the black-dotted curves serve as a guide to the eye.
 The red-solid curves are the envelope Eq. (\ref{brightsol}),
 with parameters as in (a) and (b).

\newpage

\begin{figure}[t]
\includegraphics[
width=
.7 \linewidth ]{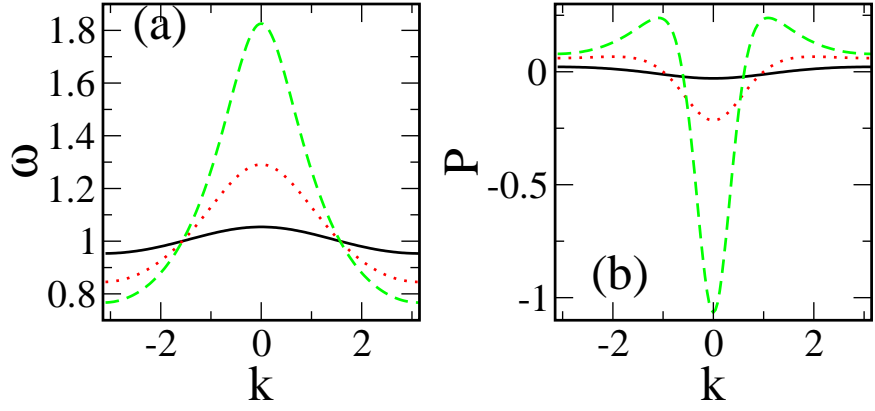}
\caption{
 (color online). (a) The linear dispersion $\omega=\omega (k)$, 
 and (b) the dispersion coefficient $P=P (k)$, 
 for $\lambda=0.05$ (black-solid curve), $0.2$ (red-dotted curve),
 $0.35$ (green-dashed curve).
}
\end{figure}

\newpage 

\begin{figure}[t]
\includegraphics[angle=0, width=.7\linewidth]{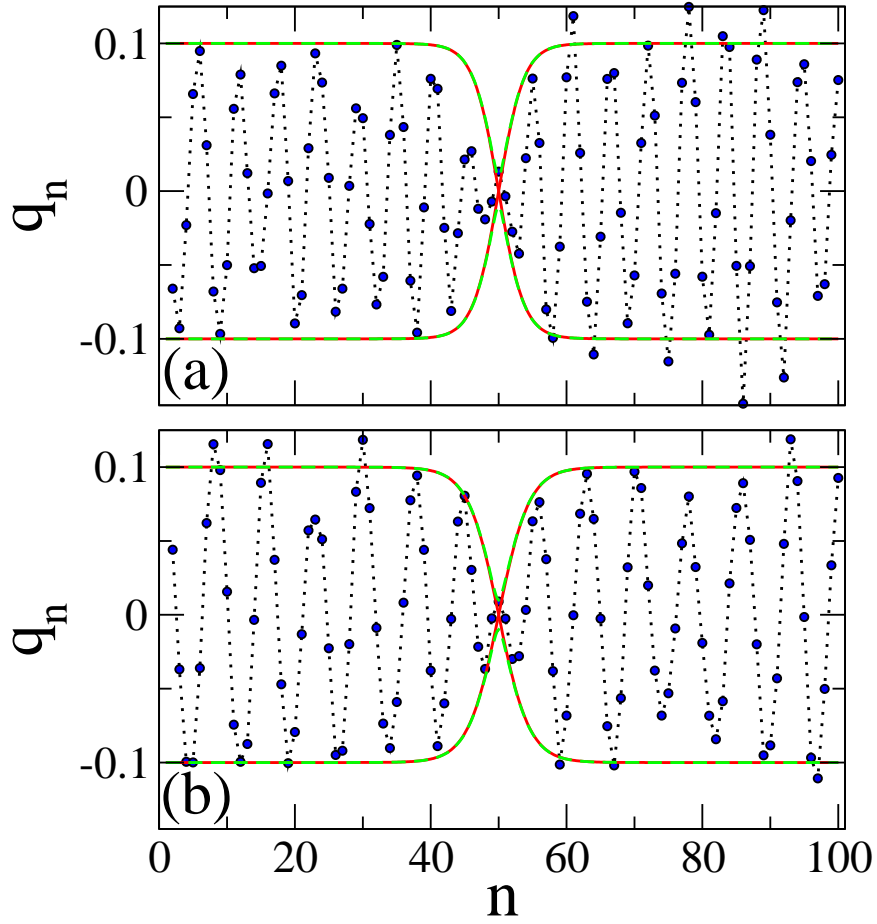}
\caption{
 (color online). Dark-type modulated wavepackets, for $\lambda=0.20$,
 $\epsilon_{\ell} = 2$, $\psi_0' =0.1$, $N=100$, and 
 (a) $\alpha=-1$, $k=1.12 > k_{cr}$ ($\omega=1.10$), $L=3.74$;  
 (b) $\alpha=+1$, $k=0.86 < k_{cr}$ ($\omega=1.16$), $L=4.51$.
 The filled circles correspond to the $q_n$'s, with $q_n$ the charge at 
 site number $n$,
 while the black-dotted curve serves as a guide to the eye.
 The red-solid curves are the envelope Eq. (\ref{darksoliton}), and the 
 green-dashed curves the envelope Eq. (\ref{greysoliton}), with parameters
 as in (a) and (b), and $d=0.995$.
}
\end{figure}

\newpage 

\begin{figure}[t]
\includegraphics[angle=0, width=.7\linewidth]{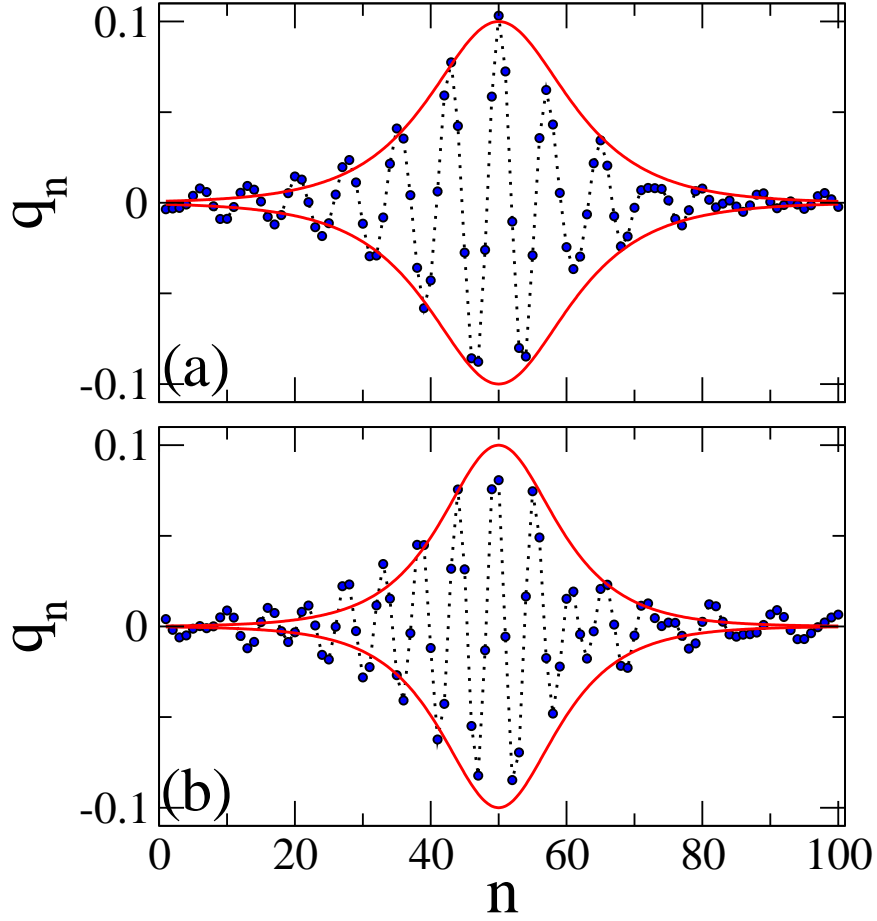}
\caption{
 (color online). Bright-type modulated wavepackets, for $\lambda=0.20$,
 $\epsilon_{\ell} = 2$, $\psi_0' =0.1$, $N=100$, and 
 (a) $\alpha=-1$, $k=0.86 < k_{cr}$ ($\omega=1.16$), $L=9.02$;
 (b) $\alpha=+1$, $k=1.12 > k_{cr}$ ($\omega=1.10$), $L=7.48$.
 The filled circles correspond to the $q_n$'s, with $q_n$ the charge at 
 site number $n$,
 while the black-dotted curves serve as a guide to the eye.
 The red-solid curves are the envelope Eq. (\ref{brightsol}),
 with parameters as in (a) and (b).
}
\end{figure}

\end{document}